\documentclass[english,twocolumn,showpacs,amssymb,prl]{revtex4}
\usepackage[T1]{fontenc}
\usepackage[latin1]{inputenc}
\usepackage{amsmath}
\usepackage{graphicx}
\usepackage{amssymb}
\usepackage{babel}

\begin{document}

\title{Interactions Between Mn$_{\text{12}}$--ac and Thin Gold Films: \\Using Mn$_{\text{12}}$--ac as Scattering Centers}

\author{Joel Means}

\altaffiliation[Currently at ]{Sandia National Labs, Albuquerque, NM 87185-1084}

\author{Winfried Teizer}

\email{teizer@tamu.edu}

\affiliation{Department of Physics, Texas A\&M University, College Station, TX
77843-4242}

\begin{abstract}
We explore the electronic interactions between a thin gold film and
a surface layer of the molecular magnet Mn$_{\text{12}}$-acetate.
Magnetoresistance measurements of the gold allow characterization
of the interactions by comparison with the theoretical predictions
of weak localization. We find that the presence of Mn$_{\text{12}}$-acetate
on the surface of the gold film leads to a reduction in elastic scattering
while increasing the spin scattering of the conduction electrons.
This is the first experimental evidence of molecular magnets being
used as scattering centers for an adjacent metallic film. 
\end{abstract}


\maketitle

\section{Introduction}

Manganese-12 acetate (Mn$_{12}$--ac) is a molecular magnet which
has been the subject of much study, both theoretically \cite{VILLAIN1994,POLITI1995,GARANIN1997,ZENG1999,GOTO2000,POHJOLA2000,TEJADA2001,LIU2002,PEDERSON2002,POSTNIKOV2006,KORTUS2007}
and experimentally \cite{CANESCHI1991,SESSOLI1993,BARBARA1995,PAULSEN1995--0,PAULSEN1995--1,NOVAK1995--0,NOVAK1995--1},
since it was first fabricated by Lis in 1980 \cite{LIS1980}. While
most of the research into the properties of Mn$_{\text{12}}$-ac has
been focused on its interesting magnetic properties, some work has
been done more recently exploring the electronic properties, in particular
the conductance through Mn$_{\text{12}}$-ac \cite{JO2006,LEUENBERGER2006,HEERSCHE2006,NI2006,ROMEIKE2006}.
Theoretical predictions have also been made concerning the electronic
structure of related Mn$_{\text{12}}$ molecules \cite{ZENG1999,VOSS2007}.
This study seeks to shed light on how the presence of these molecular
magnets on the surface of a gold film can affect the transport properties
of the conduction electrons within that gold film.

In order to explore the interactions between the Mn$_{\text{12}}$-ac
and the gold film, one must have a theoretical framework for understanding
the interactions. The theory of weak localization \cite{ABRAHAMS1979,HIKAMI1980,LEE1982,BERGMANN1983}
provides such a framework by quantifying the strengths of the various
scattering processes experienced by the conduction electrons. Conduction
electrons within a metal can undergo elastic scattering, such as surface
scattering or scattering from lattice defects; inelastic scattering,
such as electron-phonon scattering; spin scattering, such as scattering
from magnetic impurities; or spin-orbit scattering. Each of these
processes makes specific contributions to the resistance of the film
and each changes in a specific way in the presence of a magnetic field.
Weak localization predicts the change in resistance of a metallic
film in a perpendicular magnetic field, $H_{\bot}$, due to weak localization
effects to have the following dependence \cite{GIORDANO1993}:

\begin{widetext}

\begin{equation}
\frac{\Delta R\left(H_{\bot}\right)}{R}=\frac{e^{2}R_{\square}}{2\pi^{2}\hbar}\times\left[\psi\left(\frac{1}{2}+\frac{H_{1}}{H_{\bot}}\right)-\psi\left(\frac{1}{2}+\frac{H_{2}}{H_{\bot}}\right)+\frac{1}{2}\psi\left(\frac{1}{2}+\frac{H_{3}}{H_{\bot}}\right)-\frac{1}{2}\psi\left(\frac{1}{2}+\frac{H_{2}}{H_{\bot}}\right)\right]\label{eq:magnetoresistance}\end{equation}
 \end{widetext}where $R_{\square}$ is the sheet resistance of the
film and $H_{\text{n}}$ are characteristic fields given by

\begin{equation}
H_{1}=H_{0}+H_{so}+H_{s}\label{eq:H1}\end{equation}

\begin{equation}
H_{2}=\frac{4}{3}H_{so}+\frac{2}{3}H_{s}+H_{i}\label{eq:H2}\end{equation}

\begin{equation}
H_{3}=2H_{s}+H_{i}\label{eq:H3}\end{equation}
 with $H_{0}$, $H_{so}$, $H_{s}$, and $H_{i}$ being characteristic
fields corresponding to elastic, spin-orbit, spin, and inelastic scattering,
respectively. The physical quantity of interest is the scattering
time, ${\tau}_{n}$, associated with each type of scattering process.
The scattering times are related to the scattering fields by

\begin{equation}
H_{n}\tau_{n}=\frac{\hbar}{4eD}\label{eq:H_tau}\end{equation}
 where $e$ is the electron charge and $D$ is the diffusion constant
in two dimensions. Measurement of the magnetoresistance of a metal,
then, allows one to determine the scattering times associated with
the various scattering processes which the conduction electrons undergo.
By simultaneously measuring the magnetoresistance of gold films with
and without Mn$_{\text{12}}$-ac on the surface and computing how
the characteristic scattering times change, one can determine the
types of interactions taking place between the conduction electrons
in the gold and the Mn$_{\text{12}}$-ac.

\section{Experimental Setup and Results}

\begin{figure}
\begin{centering}
\includegraphics[scale=0.25]{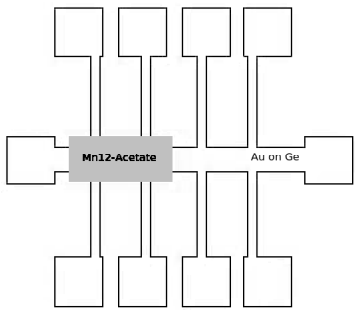} 
\par\end{centering}

\vspace{-1em}

\caption{\label{fig:Au_pattern}Schematic of pattern used for Au samples.}

\end{figure}

In order to perform the magnetoresistance measurements, a thin film
of Au was made by thermal evaporation onto a room temperature silicon
substrate with the native oxide layer and a 20$\textrm{\AA}$ Ge adhesion
layer. A shadow mask, as illustrated in figure~\ref{fig:Au_pattern},
was used to define a pattern consisting of a single current-carrying
strip with eight voltage probes, four on each side of the strip. This
setup allows a four-wire measurement of two different sections of
the strip simultaneously and with some redundancy. By placing Mn$_{\text{12}}$-ac
on half of the current carrying strip, the magnetoresistance with
and without Mn$_{\text{12}}$-ac can be measured simultaneously using
the same excitation current and magnetic field. Theoretical fits are
obtained using a least-squares fitting method to obtain the best set
of the four fitting parameters, $H_{0}$, $H_{so}$, $H_{i}$, and
$H_{s}$.

Mn$_{\text{12}}$-ac was placed on the surface of the Au film using
a simple drop-and-dry technique, similar to the established dip-and-dry
technique\cite{KIM2004,SEO2006}. A solution was made by dissolving
10~mg of Mn$_{\text{12}}$-ac powder in 10~ml of isopropyl alcohol.
Five drops of this solution were dropped onto half of the Au film
with ample drying time between drops. It was observed by AFM (see
Figure 1, ref. \onlinecite{SEO2006}) that this is sufficient to
ensure good coverage of the Au with Mn$_{\text{12}}$-ac. Samples
were then mounted in a dilution refrigerator for measurement.
Results shown are from typical samples.

\begin{figure}[h]
\begin{centering}
\includegraphics[scale=0.3]{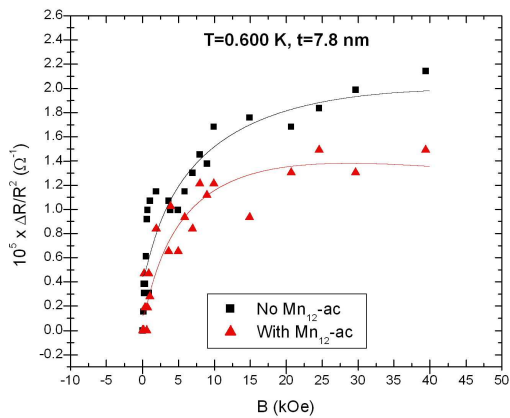} 
\par\end{centering}

\vspace{-1em}

\caption{\label{fig:7.8nm_600mK}Magnetoresistance of 7.8 nm Au film at 600
mK. Solid curves are fits to theory.}

\end{figure}

Figure~\ref{fig:7.8nm_600mK} shows the measured magnetoresistance
for a 7.8~nm Au film, with and without Mn$_{\text{12}}$-ac, at a
temperature of 600~mK. In the dilution refrigerator used in these
experiments, 600~mK was found to be an easy temperature to reach
and stabilize, while lower temperature measurements did not seem to
provide any substantial advantage. The solid curves are fits to the
theory of weak localization using Equation~\ref{eq:magnetoresistance}.
The fitting parameters, $H_{n}$, used for the fits in Figure~\ref{fig:7.8nm_600mK}
are listed in Table~\ref{tab:Characteristic-fields-78}. Also listed
are the characteristic scattering times calculated using Equation~\ref{eq:H_tau}.
The elastic scattering time is increased by approximately a factor
of two if Mn$_{\text{12}}$ is present. This indicates a decrease
in the amount of elastic scattering, consistent with a change in the
surface which reduces the amount of surface scattering. The inelastic
and spin-orbit scattering are essentially unchanged, while the spin
scattering time is reduced by around an order of magnitude. This indicates
a significant increase in the amount of spin scattering taking place.
The decrease in elastic scattering and increase in spin scattering
are consistent with a picture in which electrons which would have
scattered from the surface of the Au film are instead entering the
Mn$_{\text{12}}$-ac molecules and undergoing spin scattering, most
likely from the Mn atoms within the molecules.

\begin{table}[h]
\begin{centering}
\begin{tabular}{|c|c|c|}
\hline 
 & No Mn$_{12}$--ac & With Mn$_{12}$--ac\tabularnewline
\hline
\hline 
H$_{0}$ & 12.9$\pm$0.5 & 4.3$\pm$0.2\tabularnewline
\hline 
H$_{i}$ & 0.0145$\pm$0.005 & 0.0090$\pm$0.0045\tabularnewline
\hline 
H$_{so}$ & 0.97$\pm$0.02 & 0.58$\pm$0.02\tabularnewline
\hline 
H$_{s}$ & 0.0040$\pm$0.0025 & 0.0080$\pm$0.0020\tabularnewline
\hline 
$\tau_{0}$ & (3.62$\pm$0.07)$\times$10$^{-15}$ & (6.31$\pm$0.15)$\times$10$^{-15}$\tabularnewline
\hline 
$\tau_{i}$ & (3.66$\pm$1.20)$\times$10$^{-12}$ & (3.97$\pm$1.92)$\times$10$^{-12}$\tabularnewline
\hline 
$\tau_{so}$ & (4.85$\pm$0.01)$\times$10$^{-14}$ & (4.76$\pm$0.05)$\times$10$^{-14}$\tabularnewline
\hline 
$\tau_{s}$ & (1.91$\pm$1.17)$\times$10$^{-11}$ & (3.59$\pm$0.82)$\times$10$^{-12}$\tabularnewline
\hline
\end{tabular}
\par\end{centering}

\vspace{2em}

\caption{\label{tab:Characteristic-fields-78}Characteristic fields and scattering
times for 7.8 nm gold film at 600 mK.}

\end{table}

\begin{figure}[h]
\begin{centering}
\includegraphics[scale=0.3]{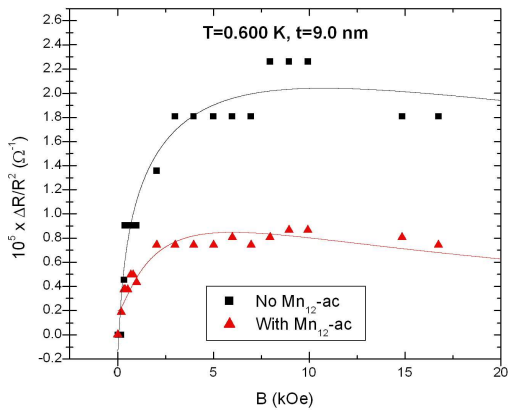} 
\par\end{centering}

\vspace{-1em}

\caption{\label{fig:9.0nm_600mK}Magnetoresistance of 9.0 nm Au film at 600
mK. Solid curves are fits to theory.}

\end{figure}

Figure~\ref{fig:9.0nm_600mK} shows the magnetoresistance for a 9.0~nm
Au film, also at 600~mK. Fitting parameters used for the solid curves
are listed in Table~\ref{tab:Characteristic-fields-90}. As with
the thinner sample, the inelastic and spin-orbit scattering times
remain unchanged within experimental error. It should be noted that
due to the small changes being measured, we observe digitization of
the data, indicating that the measurements are close to the experimental
resolution. The changes in resistance were on the order of the smallest
digit which could be resolved with the experimental setup used. This
digitization also leads to larger relative errors in the numbers reported.
Again, there is a significant increase in the elastic scattering times,
by a factor of approximately four. There is also a decrease in the
spin scattering time by about an order of magnitude. This behavior
is consistent with that seen for the thinner sample.

\begin{table}[h]
\begin{centering}
\begin{tabular}{|c|c|c|}
\hline 
 & No Mn$_{12}$--ac & With Mn$_{12}$--ac\tabularnewline
\hline
\hline 
H$_{0}$ & 3.1$\pm$0.4 & 0.5$\pm$0.1\tabularnewline
\hline 
H$_{i}$ & 0.0015$\pm$0.001 & 0.0050$\pm$0.0045\tabularnewline
\hline 
H$_{so}$ & 0.23$\pm$0.02 & 0.12$\pm$0.02\tabularnewline
\hline 
H$_{s}$ & 0.0005$\pm$0.0004 & 0.0050$\pm$0.0030\tabularnewline
\hline 
$\tau_{0}$ & (7.47$\pm$0.48)$\times$10$^{-15}$ & (1.88$\pm$0.18)$\times$10$^{-15}$\tabularnewline
\hline 
$\tau_{i}$ & (2.63$\pm$1.66)$\times$10$^{-11}$ & (8.78$\pm$7.72)$\times$10$^{-12}$\tabularnewline
\hline 
$\tau_{so}$ & (9.99$\pm$0.21)$\times$10$^{-14}$ & (7.75$\pm$0.51)$\times$10$^{-14}$\tabularnewline
\hline 
$\tau_{s}$ & (4.60$\pm$2.16)$\times$10$^{-11}$ & (2.70$\pm$1.43)$\times$10$^{-12}$\tabularnewline
\hline
\end{tabular}
\par\end{centering}

\vspace{2em}

\caption{\label{tab:Characteristic-fields-90}Characteristic fields and scattering
times for 9.0 nm gold film at 600 mK.}

\end{table}

\begin{figure}[h]
\begin{centering}
\includegraphics[scale=0.3]{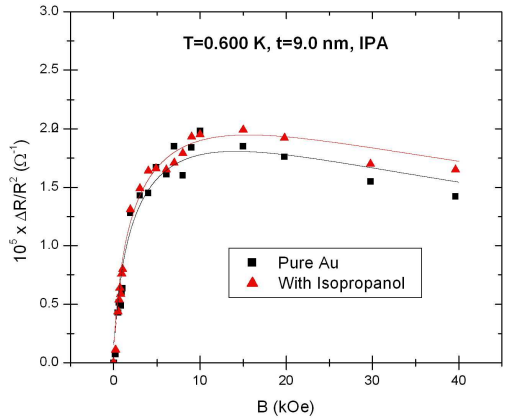} 
\par\end{centering}

\vspace{-1em}

\caption{\label{fig:9.0nm_600mK_isopropyl}Magnetoresistance of 9.0 nm Au film
at 600 mK with isopropyl alcohol. Solid curves are fits to theory.}

\end{figure}

As a control experiment, a gold sample was prepared with pure isopropyl
alcohol on half of the current carrying path in order to eliminate
the possibility of the effect being due to residual contaminations
in the isopropyl alcohol. The experimental data, along with theoretical
fits, are shown in Figure~\ref{fig:9.0nm_600mK_isopropyl}. The fitting
parameters which provided the best fit are listed in table~\ref{tab:Char-fields-9.0A-600mK_IPA}.
Within experimental error, there are no significant changes in any
of the scattering times due to the presence of the isopropyl alcohol
or to any contaminations within the isopropyl alcohol. This indicates
that the changes seen in the samples with Mn$_{\text{12}}$-ac are
due to the Mn$_{\text{12}}$-ac and not to the isopropyl alcohol which
was used as a solvent for sample preparation.

\begin{table}[h]
\begin{centering}
\begin{tabular}{|c|c|c|}
\hline 
 & Pure Au & With Isopropyl Alcohol\tabularnewline
\hline
\hline 
H$_{0}$ & 3.2$\pm$0.2 & 3.6$\pm$0.4\tabularnewline
\hline 
H$_{i}$ & 0.0035$\pm$0.0015 & 0.0015$\pm$0.0010\tabularnewline
\hline 
H$_{so}$ & 0.29$\pm$0.02 & 0.30$\pm$0.02\tabularnewline
\hline 
H$_{s}$ & 0.0010$\pm$0.0005 & 0.0010$\pm$0.0005\tabularnewline
\hline 
$\tau_{0}$ & (7.30$\pm$0.25)$\times$10$^{-15}$ & (6.92$\pm$0.38)$\times$10$^{-15}$\tabularnewline
\hline 
$\tau_{i}$ & (6.67$\pm$1.85)$\times$10$^{-12}$ & (2.86$\pm$1.81)$\times$10$^{-11}$\tabularnewline
\hline 
$\tau_{so}$ & (8.05$\pm$0.34)$\times$10$^{-14}$ & (8.25$\pm$0.09)$\times$10$^{-14}$\tabularnewline
\hline 
$\tau_{s}$ & (2.33$\pm$1.40)$\times$10$^{-11}$ & (3.21$\pm$1.46)$\times$10$^{-11}$\tabularnewline
\hline
\end{tabular}
\par\end{centering}

\vspace{2em}

\caption{\label{tab:Char-fields-9.0A-600mK_IPA}Characteristic fields and scattering
times for 9.0 nm gold film at 600 mK with isopropyl alcohol.}

\end{table}

\section{Conclusion}

It has been shown that the presence on Mn$_{\text{12}}$-ac on the
surface of a thin Au film causes significant changes in the magnetoresistance
of the Au film. Fitting of the experimental data to the predictions
of the theory of weak localization allows characterization of the
types of changes taking place. In particular, there is an enhancement
of the spin scattering of the conduction electrons, coupled with a
reduction in the elastic scattering. This is consistent with electrons
entering the Mn$_{\text{12}}$-ac and undergoing spin scattering,
rather than simply scattering from the surface of the Au film. This
is the first time that experimental evidence has been seen indicating
that a surface layer of molecular magnets can be used as scattering
centers to change the electron transport properties of a metallic
film.

\end{document}